# BASIC CONCEPT IN PLASMA DIAGNOSTICS

V. N. Rai

Laser Plasma Division

Raja Ramanna Centre for Advanced Technology

Indore-452 013

## ABSTRACT

This paper presents the basic concept of various plasma diagnostics used for the study of plasma characteristics in different plasma experiments ranging from low temperature to high energy density plasma.

## 1. INTRODUCTION

It is well known that 99% of the matter in the universe is in the form of plasma, which is the fourth state of matter after solid, liquid and gas. The natural sources of plasma are the atmosphere, sun, stars and gaseous nebula. However various types of plasma are generated in laboratories for different applications, which include gaseous discharge, arcs, laser produced plasma as well as tokamak plasma. These plasma sources have various applications in different fields of research and industry. This is the reason that during the past few decades plasma physics has been established as a major research field. The major driving force behind the research on the plasmas has been and still continuing to be is the better prospect of generating economically significant amount of power from controlled thermonuclear fusion. The two main technique being persued for fusion is Inertial Confinement Fusion (ICF) based on laser produced plasma and the other one is Magnetically Confined Fusion (MCF) based on gaseous discharge. On the other hand most of the other applications use a cold plasma having temperature around few eV in comparison to high temperatures encountered in fusion grade plasma. Even the density range is also quite different. This categorizes plasma in two types as cold and hot plasma, which compelled plasma physicist to develop different kinds of diagnostics for measuring accurately the parameters of plasma for achieving the goals as well as to better understand the physics of plasma. This involves the knowledge of a wide variety of branches both theoretical as well as experimental like the other sub-discipline of physics.

The techniques developed for diagnosing the properties of plasma is known as plasma diagnostics.

The main objective of the plasma diagnostics is to deduce informations about the state of the plasma from practical observations of physical processes and their effects. Accurate and reliable measurements of the plasma condition in the various plasma experiments are an important objective for making a significant progress in the field. Since the plasma behavior is dependent on the condition in which it is produced, so it demands knowledge of many plasma quantities for its description as well as for its comparison with relevant theory. Particularly in fusion experiments many important quantities are either measured with poor accuracy or even not measured at all. Plasma research and particularly fusion research cover a wide area of basic physics and use most advanced technology.

The plasma diagnostics can be cetegorized in various ways. First of all one can consider cold and hot plasma. Material probes such as the most common Langmuir probe can be inserted in the cold plasma. However in the hot plasma ($T_e \geq$ few keV) the use of material probes is limited to the extreme edge of the plasma ($T_e \leq 50eV$) and the non − invasive techniques must be used to diagnose other plasma regions. These includes passive methods, which detects radiation or particles emitted spontaneously by the plasma, where as in active methods radiations or particles produced by external sources are used to probe the plasma. Another way of grouping them are on the basis of measured plasma parameters such as plasma density and temperature etc. But one can see that many diagnostics measure more than one parameter, so this classification will lead to a repetition in the diagnostics. This article discusses basic concept of some of the important plasma diagnostics being used for the study of cold as well as hot plasmas produced in the laboratory including the fusion plasmas.

## 2. ELECTRIC AND MAGNETIC PROBE

### 2.1 Langmuir Probe

The most basic probe used for measuring the properties of plasma is an electrostatic probe. This technique was developed by Langmuir in 1924. It is simply a small metallic electrode in the form of wire inserted into the plasma. A variable biasing voltage is applied on this electrode, which may be positive or negative with respect to the plasma. The current collected by the probe with respect to the applied voltage provides various informations about the plasma condition. It has been observed that under certain conditions, such as in the presence of strong magnetic field as well as in the turbulent plasma conditions these measurements become difficult. However inspite of all difficulties, it can provide information about the local parameters (plasma density and temperature) in comparison to the other plasma diagnostics in the cold plasma.

An important fact about the probe is that it is a very simple device but has very complicated theory. This is because probe acts as a boundary and the equations governing the plasma motion change its characteristics. Due to the quasi-neutral behaviour of plasma a sheath is formed around the probe, which can sustain a large electric field. The characteristics of the Langmuir probe can be easily understood by plotting the I-V curve. Here I is the current flowing in the probe when a biasing voltage V is applied to it with respect to the plasma or the body of the vessel containing plasma. This curve can be plotted easily in a steady state discharge, whereas it can be obtained point by point in pulsed discharge or by fast sweeping of the bias voltage in few microseconds. Biasing voltage is changed from negative to positive voltage. The probe characteristics have two types of saturation current, (1) electron saturation current when the probe is positively biased and (2) ion saturation current in the case of negative biasing. The amount of biasing voltage, for which the probe current is zero, is known as floating potential. Probe current increases in between the ion and electron saturation current due to Maxwellian distribution in the plasma. When the probe potential is increased towards positive only those particles (electrons) are collected on the probe, which has similar energy, rest of the high-energy particles will not be collected. Similar is the condition after further increasing the probe current. When potential becomes equal to the plasma temperature nearly all the electrons are collected, which resulted in electron saturation current. Infact, perfect electron saturation currents are not observed due to the presence of high-energy electrons in the tail of Maxwellian distribution in the plasma. Turning point near the electron saturation current is known as space charge potential. The probe current can simply be given by

$$I = I_0 \, e^{\,eV/kT} \qquad\qquad \text{-------------------------------(1)}$$

After taking the log one can write it as

$$Ln\ I = Ln\ I_0 + eV/kT \qquad \text{--------------------------------(2)}$$

This shows that plot of LnI-V curve can provide information about the plasma temperature from the slope of the straight line. Since probe current mainly depend on electron current it can provide better information about electron temperature in comparison to ion temperature. It has been observed that inspite of a difficult probe theory for various plasma conditions; a simple relation can approximate probe current in the absence of disturbance in the plasma as

$$J = nev \qquad \text{----------------------------------(3)}$$

Where J is the current density on the probe. n is the plasma density and v is the velocity of the electron corresponding to plasma temperature. If some type of instability is present in the plasma one can get information about it by measuring the fluctuation in ion saturation current. Finally it is possible to obtain information about local plasma density, electron temperature and space potential along with density fluctuation, if a probe is placed in the plasma without disturbing it.

## 2.2  Double Probe

Double Langmuir probes can be used in plasma, where reference point is not available. The example of this type of plasma is RF discharge, which is an electrode less plasma. Double probe (electrode) is used simultaneously in the plasma. The biasing voltages on the probe are varied from negative to positive potential for getting an I-V characteristic of the probe. This provides ion saturation current in the negative ($I_{1+}$) as well as in positive direction ($I_{2+}$). This curve passes through the origin (V= 0, I=0), where current is linearly varying with applied voltage. The rate of change of current with respect to voltage in the linear range can be expressed as

$$\frac{dI}{dV} = \frac{e}{kT_e} \frac{I_{1+} \cdot I_{2+}}{I_{1+} + I_{2+}} \qquad \text{--------------------------------(4)}$$

This shows that once the plasma temperature ($kT_e$) is known plasma density can be estimated from the amplitude of ion saturation current.

As discussed earlier probe measurement has several limitations based on its complicated theory. Particularly in the case of collisional plasma or in the presence of magnetic field, probe measurements need correction for getting meaningful results.

## 2.3 Magnetic Probe

In some cases strong currents are present in the plasma, particularly in the tokamak plasma. Presence of plasma current generates a magnetic field around the plasma current. Plasma current in the magnetized plasma generates different types of magneto hydrodynamic (MHD) instability, which is observed as a fluctuation superimposed on the magnetic field generated due to plasma current. Presence of a search coil (small area solenoid having only few turns) near the plasma current can measure any fluctuation in the magnetic field. An analysis of voltage induced in the search coil provides information about the MHD instability present in the plasma.

If the search coil has n number of turns on a small area a, then the induced voltage in the search coil can be given by

$$V = n\frac{d\phi}{dt} = na\frac{d(\delta B)}{dt}$$ -------------------------------(5)

where $\delta B$ is the small fluctuation in the magnetic field around the plasma. This provides a differential signal, which should be integrated for getting direct information about the fluctuation in magnetic field. If RC is the time constant of an integrator, then amplitude of the fluctuating magnetic field can be obtained from the induced voltage on the search coil and can be written as

$$\delta B = RC\ \frac{V}{an}$$ ------------------------------(6)

Before using the search coil for measurement, it should be properly shielded. Another important point is that search coil should not be in contact with the hot plasma. It is useful particularly for measuring either local magnetic field or its fluctuations in the tokamak or in laser produced plasma.

## 2.4 Rogowski Coil

It is possible to measure the amplitude of current flowing in the plasma by integrating the induced magnetic field around the plasma column using a long solenoid popularly known as Rogowski coil. Amplitude of the total current can be given by

$$I = \int B.dl \qquad \text{-------------------------------(7)}$$

The plasma current will induce a voltage in Rogowski coil, which will be proportional to the rate of change of plasma current with time. Finally expression for the induced voltage in Rogowski coil in term of plasma current can be given by

$$V = \frac{\mu_0 NA}{2\pi a} \cdot \frac{dI_P}{dt} \qquad \text{------------------------------(8)}$$

Where $I_p$ is the plasma current, a is the radius of plasma column, N is the number of turns in the Rogowski coil and A is its area of cross section. After integrating the Rogowski coil signal by a suitable active (electronic) or passive integrator one can get information about the time evolution of the plasma current. Final expression for the plasma current can be given by

$$I_p = \frac{2\pi a}{\mu_0 NA} \cdot RC.V \qquad \text{-----------------------------(9)}$$

Here RC is the time constant of the integrator. For better integration of Rogowski coil signal, integration time must be nearly 10 times more than the total duration of the plasma current. In this case measured plasma current remains independent of position of the plasma column inside the Rogowski coil. To avoid any electrostatic pick up on the signal Rogowski coil must be shielded properly. Coil winding must return from the end of initial winding (bifilar winding), other wise the presence of any other magnetic field passing in the same or opposite direction of the plasma current can also generate an electrical signal considering Rogowski solenoid as one turn coil. These factors are important for designing a Rogowski coil to provide nearly true shape of plasma current in time. This is an useful diagnostics for a tokamak machine.

## 3. CHARGED PARTICLE COLLECTION

It is well known that plasma is a collection of charge particles such as ions and electrons as well as neutral particles. The energy of these particles is mainly dependent on the temperature of the plasma. Since the plasma has Maxwellian distribution. It contains charged particles of wide energy spectrum. There are two important processes in the plasma that generates high-energy ions and electrons in the plasma. The presence of strong electrostatic field in the plasma as a result of ambipolar diffusion creates high-energy particles in the plasma due to the acceleration process. However, in the laser produced plasma main source of high energy particles (suprathermal electrons) are the resonance absorption of laser light in the plasma near the critical density surface, where frequency of the laser becomes equal to the plasma frequency.

## 3.1   Methods of Energy Analysis

There are two main processes, which are used to measure the energy spectrum of the charged particles in the plasma. The simplest method is to use time of flight measurement. In this case a simple particle detector is placed at a certain distance (d) from the plasma source and time of arrival of charged particles on the detector ($\tau$) are recorded. This provides information about the energy of the charge particle as is shown by the expression.

$$E = \frac{1}{2} m \left( \frac{d}{\tau} \right)^2 \qquad \text{-----------------------------------(10)}$$

where m is the mass of the charged particle, d is the distance of plasma source from the particle detector and $\tau$ is the time of flight. This technique is mainly useful in the pulsed plasma source particularly the laser-produced plasma. Various types of detectors are used for this purpose such as retarding grid analyzer, Faraday cup, biased collector, electron multipliers, scintillator photomultiplier combination as well as solid-state detectors. Scintillator, photo multiplier combination is an important diagnostics for the time of flight measurement of fusion neutrons, which will be discussed later on.

Retarding grid analyzer is the simplest system used for the measurement of energy distribution of ions and electron in the plasma. It consists of a circular disc for collecting the charge particles. A grid (fine metallic mesh) is used before collector that is biased in order to reject the charge particles of certain energies. If biasing voltage is V then all the particles having energy below this will be rejected and rest will be collected by the collector. By changing the biasing voltage on the grid one can get an I-V curve, which shows decrease in collector current (I) by increasing the biasing voltage. In the case of variable positive potential on grid, it will work as an ion energy analyzer, where as for variable negative biasing it works as electron energy analyzer. From the I-V characteristics it is possible to find the plasma temperature also. The temporal evolution of the collector current I (t) can be given by

$$I(t) = Q.\frac{dN}{dt} \qquad \text{-----------------------------(11)}$$

Where Q is the charge on the plasma particle and dN/dt is the rate of change in the number of collected particles on the disc. Using eq. (10) and (11) one can write expression for total number of particles as well as total energy collected on the disc as

$$N_{Total} = \int_0^\infty \frac{I(t)}{Q}.dt \qquad \text{----------------------------(12)}$$

An expression for total energy also can be written as

$$E_{Total} = \int_0^\infty \frac{1}{2}\left(\frac{d}{\tau}\right)^2.I(t).dt.\frac{m}{Q} \qquad \text{---------------------------(13)}$$

It is important that spurious signal due to secondary emission from grid should be minimized. In some cases simple disc collector is replaced by a Faraday cup to make the collection angle small.

Another technique for measuring the energy spectrum is based on the deviation of charged particles from their path in the effect of magnetic / electric field or in the combination of both. Amplitude of deviation from the initial path depends mainly on the energy of the charged particles. Mass spectrometer is also used to obtain the energy spectrum along with Q/m information about the charged particles. In this case a drift tube is used to separate the incident ion velocity by time of flight method, whereas electrostatic technique is used to separate various Q/m state in a given velocity component. Charge particles of different Q/m are selected by applying electrostatic potential. The particles of different Q/m are collected at different locations with the help of array detectors. In this case signal at each detector is proportional to n, v and Q/m. Finally it provides energy spectrum of each Q/m state.

## 4. INTERFEROMETRY

In many plasmas (particularly high temperature) it is not possible to use material probes to determine internal plasma parameter such as plasma density. So a non-perturbing method is required as diagnostics. In such cases electromagnetic wave are used for probing the plasma provided there intensity is not to high and such wave causes negligible perturbation to the plasma. This technique uses the modification in the free space propagation of electromagnetic wave due to electrical properties of the plasma. Interferometry is the main experimental technique used for measuring the plasma refractive index properties. Interferometer uses a monochromatic electromagnetic radiation as a probe source, which is splitted in two parts. After traveling equal optical length both the waves are recombined to provide an interference pattern. If some other medium such as plasma is created in one of the path of electromagnetic wave then a path difference is created in the two arms, which results in a distortion in the interference pattern obtained in the absence of plasma. A proper analysis of phase shift from initial fringe can provide information about the plasma density present in the path of electromagnetic wave. For this one has to consider the effect of refractive index of plasma on the interference pattern.

Generally plasma is considered as a dielectric medium, which has its refractive index ($\mu$) dependent on the plasma frequency $\omega_p$ ($\omega_p = \dfrac{4\pi n e^2}{m_e}$, where n, $m_e$ and e are the density, mass and charge of electrons). If $\omega_L$ is the frequency of electromagnetic wave (may be a laser or microwave) then refractive index of plasma can be written as

$$\mu = \left(1 - \frac{\omega_P^2}{\omega_L^2}\right)^{\frac{1}{2}} \cong 1 - \frac{1}{2}\frac{\omega_P^2}{\omega_L^2} \quad \text{-------------------------------(14)}$$

Presence of plasma in one of the path of an electromagnetic wave can generate a path difference, which will be dependent on the refractive index as well as on the dimension of the plasma (L). This path difference can be written as

$$(1 - \mu)L \cong \frac{2\pi n e^2 L}{m_e \omega_L^2} \quad \text{------------------------------(15)}$$

The phase difference corresponding to this path difference can be given by

$$\Delta\phi \cong \frac{2\pi(1-\mu)L}{\lambda_L} \quad \cong \frac{ne^2 L\lambda_L}{m_e c^2} \quad \text{-----------------------------(16)}$$

Here $\lambda_L$ is the wavelength of probe beam. The fringe separation $\Delta x$ on an interferogram equals phase difference of $2\pi$. This way experimentally measured small shift $\delta$ in the fringe due to the presence of plasma can provide information about the plasma density. An expression for the plasma density can be written in the term of fringe shift as

$$n = \frac{2\pi mc^2\delta}{\lambda_L e^2 L\Delta x} \quad \text{-------------------------(17)}$$

The probing electromagnetic wave (Laser or microwave) provides integrated information about the plasma density. Therefore information about the density profile can be obtained by an Abel inversion. The choice of probing beam depends on density range to be probed. Normally for probing the high-density plasma lower wavelength probe is needed. This is the reason why infrared to microwaves are used for interferometry in tokamak depending on the minor radius of tokamak and green to ultra violet wavelengths are suitable for measuring density of laser-produced plasma.

Interferometric probe beam can also provide information about the magnetic field present in the plasma particularly in the case of tokamak plasma. This is possible by measuring the change in the angle of polarization of the probe beam in the presence of magnetic field. A change in the angle of polarization $\phi$ can be expressed as

$$\phi = 2.6\times10^{-25}\lambda_L^2\int n(r)B(r)dr \quad \text{-----------------------------(18)}$$

In the case of laser produced plasma wavelength of the probe beam must be smaller than the main laser beam creating the plasma. Because lower wavelength laser can penetrate to a higher density level. Reflectometry has also been used to find the density of the plasma with a consideration that a probe beam of certain frequency cannot penetrate the critical density of the plasma. Particularly this technique has been applied in the tokamak plasma using microwave as a probe beam.

## 5. PLASMA SPECTROSCOPY

Spectroscopy is probably the most powerful diagnostic technique available to plasma physicists. Plasma spectroscopy is the study of electromagnetic radiation emitted by the ionized media. However, in contrast to the conventional spectroscopy, where one is only interested in the atomic structure of an isolated atom, radiation from a plasma depends not only on the properties of the isolated radiating species, but also on the properties of the plasma in the immediate environment of the radiator. This dependence on the plasma properties is a consequence of the fact that ions and electrons interact with the radiating species through the process of ionization, recombination, excitation and de-excitation, thus determining the radiator state. However long range coulomb potential also causes purturbation. With the help of this technique large number of plasma parameters can be obtained such as impurity concentration, effective charge ($Z_{eff}$) of the plasma, radiative power losses, electron temperature $T_e$, electron density $n_e$ as well as the level by which the plasma charge distribution departs from the ionization equilibrium. Plasma spectroscopy has been developed to investigate high temperature laboratory plasma due to particular interest in the controlled thermonuclear fusion experiments. Tokamak plasmas have temperature profiles with a central electron temperature of several keV and a boundary temperature in the eV range. As a consequence, impurity atoms are multiply ionized or fully stripped depending on their location in the plasma. Light impurities (O, C, N, Cl) are completely ionized over most of the plasma region and radiate predominantly at the plasma periphery. Heavy impurities (Cr, Fe, Ni) are only partially ionized over the entire plasma and radiate strongly, which is why a contamination of only 1% iron makes the ignition (fusion) impossible. Generally the energy of the emitted photons is $h\nu = kT$, that is, the main part of the emitted radiation is in the VUV and soft X-ray spectral regions (T = 1 keV associated with $\lambda = 1.2 \times 10^{-9}$ m).

A complete evaluation of impurity concentrations and radiative losses in a given plasma would require extensive spatially resolved spectroscopic measurements of at least the strongest transitions of all ionization stages for each important impurity element. The measurement of strong emission lines from ions and a model has to be used to deduce the complete impurity properties. For this type of measurements, spectroscopic instruments are needed, which covers wavelength range from visible to X-ray region and is calibrated for absolute intensity. Normally gratings are used in the visible and UV regions as dispersive elements with high efficiency and high resolving power of order $10^4$. Window and mirrors are available only down to about 1200 A$^0$. For wavelength between 300 $A^0$ and 20 $A^0$ the grazing incidence technique is applied. The resolving power decreases with wavelength (down to $10^3$ at 20 $A^0$). In the X-ray regime below 20 $A^0$, Bragg reflection on crystals replaces ruled gratings. The resolving power depends on the properties of the crystal and can be $> 10^4$. X-ray transmission grating spectrograph (resolving power $\sim 10^2$, sensitive for few $A^0$ to few 100 $A^0$) has been proved to be a good tool to find the radiation temperature inside the laser irradiated gold cavity along with the spatial and absolute spectral information. These gold cavity or Hohlraums are used for indirectly driven inertially confined fusion experiments, where X-rays (black body radiations) are used to implode the fusion pellet (D -T filled micro balloons). X-ray back lighting experiments also provide various important informations about the plasma.

A variety of photoelectric detectors are available to record the spectrum electronically. An open photomultipliers, channeltron, microchannel plates and intensified CCD camera cover the whole range from the grazing incidence to visible regime. At shorter wavelength the quantum efficiency is poor and proportional counters are a better choice. Light conversion by scintillator-photomultiplier system is less popular on tokamaks, because of their high sensitivity to hard X-rays radiation and magnetic stray fields.

With the help of spectroscopic measurements, one can get information about the influx of neutral impurity and hydrogen atoms entering the plasma, which are ionized at the periphery. With the help of suitable computer codes impurity concentration and total radiation loss also can be inferred. Here the Doppler broadening of impurity lines is a convenient method to measure the ion temperature of the plasma provided the impurity temperature is sufficiently closely coupled to the hydrogen ion temperature. Various other informations also can be obtained from spectroscopic measurements of fusion plasma.

## 6. PLASMA IMAGING

### 6.1 X-Ray Pin Hole Camera

X-ray pinhole camera is used to image the x-ray emitting plume of laser-produced plasma onto an image plane. Functioning of pinhole camera follows the principle of geometrical optics. The image becomes more distinct as the pinhole is made smaller. But it should not be so small that diffraction becomes important and affects the spatial resolution of the camera. For a small pinhole, Raleigh criterion are used where a point forms a diffraction pattern in such a way that first minimum of one pattern coincides with the first minimum of second. The point will be at the limit of resolution θ defined by

$$\theta = 1.22 \; \lambda/a \qquad \text{-------------------------------(19)}$$

In simple term spatial resolution can be given as

$$\Delta s = a \; [(M+1)/M] \qquad \text{------------------------------- (20)}$$

Where M is the magnification of image. Δs in fact is nearly equal to the pinhole size (a). X-ray pinhole camera is one of the important diagnostics used to study the symmetry of implosion in direct and indirect drive inertial confinement fusion experiment. It provides time integrated spatial information about the X-ray emission from the planar or micro balloon targets plasma plume.

## 6.2 Picosecond Time Resolution X-Ray and Optical Streak Camera

Streak camera is one of the most versatile instruments used today for high-speed photometry in the field of physics, chemistry, biology and non-linear optics. Particularly rapidly time varying hydrodynamic and radiative processes in laser produced plasma and inertial confinement fusion experiments demand detailed studies of time resolved x-ray and optical emission from laser produced plasma. The conventional optical detectors such as biplanar photodiode and PIN silicon photodiode used alongwith the fastest available oscilloscope provide a time resolution of ~$10^{-10}$ s.   Various other techniques generally used for investigating very fast optical events are autocorrelation method, second harmonic generation, two-photon fluorescence and four-photon parametric mixing. However, certain amounts of   errors   (inaccuracies) are always associated with these measurements.  Amongst the entire available methods streak photography is considered to be the best for recording ultrafast optical phenomenon with an excellent time resolution (<1ps).   It can provide temporal profile of any optical event directly even in a single shot operation.    In addition to the time duration, rise time as well as shape of the optical pulse also can be recorded along with intensity modulation  (fine structure) if present during the optical event.    However measurements will be limited by the time resolution of the streak camera.    Streak cameras with nanosecond time resolution were built long back, but the temporal resolution of picosecond or sub picosecond order has been achieved only a few years back. New streak camera with femtosecond time resolution has also been reported. Highly sensitive streak cameras with Pico second time resolution are now commercially available, which are capable of recording any phenomenon in the spectral range extending from near infrared to ultraviolet and soft X-ray (100 eV - 10 keV) regions with the help of a suitable photocathode attachment. In fact the streak camera can functionally be referred to as a > 300 GHz bandwidth optical oscilloscope. However, the streak camera possesses many more special features, which scores over an ordinary oscilloscope, for example, it can provide three dimensional informations such as spatially time resolved as well as time resolved spectroscopic intensity streak camera attached with dispersing elements) profiles. Either a photographic system or a digital image memory (CCD camera) are used to store and analyze three dimensional streak images, whereas later one provides a nearly real time measurements of the events.

## 6.3 X-RAY FRAMING CAMERA

The x-ray framing camera has been useful in capturing several images from a single optical event gated at an adjustable inter frame time. The main application of gated x-ray detectors has been in the field of inertial confinement fusion and X-ray laser research. The shutter speed of the framing camera has been decreased to < 50 ps which is needful to avoid motional blurring during the exposure of moving and X-ray emitting objects. The plasma expansion velocity is found typically $\sim 10^7$ cm/s in laser produced plasma and inertial confinement fusion experiment. If spatial resolution of $\sim 10$ µm is required then shutter time speed of < 100 ps is needed to freeze the motion. There are two methods for gating the images such as gating of X-ray image converter tube, which is nothing but the streak tube. The second process is the gating of proximity focused microchannel plate detector. In the first case where image converter tube is used for framing the camera two main changes are required in comparison to the X-ray streak camera. A small gate pulse of 200 volts and of 3ns time duration is applied on the photocathode at a time interval of 5ns, which allows the photoelectrons to pass through the deflection plates. These photoelectrons make an image on the phosphor screen after a proper deflection. A staircase type of voltage are applied on the deflection plate in stead of a fast rising pulse as is required in the case of streak camera. During the rising part of staircase voltage the location of image shifts on the phosphor screen where as during the flat portion image remains still on the phosphor screen and was recorded. The main problem with this technique is the loss of spatial resolution at short exposure time, mainly in the direction of deflection. It also has limitation in providing large number of frames with better time and space resolution.

Nowadays, a second technique is being used which has overcome all the shortcoming of the previous techniques. It consists of a microchannel plate coupled with a phosphor screen. There are gold-coated low impedance strip transmission lines on the front surface of MCP where as the opposite side has a plane-conducting surface at a ground potential. The gold coated transmission line has three design out of which S shape transmission line on MCP front are mainly used due to less complexity in its electronics. A very short (< 100ps) time duration pulse propagates along this transmission line and activates different locations of transmission line at a certain time interval depending on the propagation velocity of pulse along the transmission line. The X-ray emitting source is imaged on this transmission line using an array of multiple pinholes in between source and MCP. In this process more than 10 images can be captured at a time separation decided by the spatial separation of images on the MCP and the speed of pulse propagation through the transmission line.

## 7. DETECTION OF NEUTRON AND CHARGED FUSION PRODUCT

The particles emitted from the fusion reaction have been used as a diagnostic tool. Fusion reaction takes place at higher plasma temperature as the fusion reaction rate increases considerably. The most important fusion reaction are given below

1.  $D + D \rightarrow$  T     (1.011MeV) + p (3.022 MeV)          ; 50%

$^{3}$He (0.820MeV) + n (2.449 MeV)          ; 50%

$^{4}$He          + $\gamma$ (23.8 MeV)          ; $10^{-5}$%

2.  $T + D \rightarrow$  $^{4}$He  (3.560 MeV) + n  (14.029 MeV)          ;100%

$^{5}$He          + $\gamma$ (16.6 MeV)          ;$7.10^{-3}$%

3.     $^3$He + D → $^4$He (3.713 MeV) + p (14.640 MeV)          ; 100%

        $^5$Li              + γ (16.5 MeV)              ;3.10$^{-3}$%

4.     T + T     →   $^4$He (1.260 MeV) + 2n (11.327 MeV)          ; 100%

Here D, T, p and n signifies deuterium, tritium, proton and neutron. To study the fusion reactions, various neutron diagnostics have been developed in order to detect the of 2.5 MeV neutrons from deuterium plasma and 14 MeV neutrons from tritium plasma. Neutron yield from the fusion plasma is dominated by the hot central core of the plasma and the central ion temperature can therefore be obtained from it. The neutron yield is very sensitive to distortion in the ion distribution from a Maxwellian, especially in the high-energy tail. A more direct measurement of the ion temperature is obtained by neutron spectrometry. The energy distribution of the neutrons is largely determined by the reaction energy, however a smaller contribution is due to the ion motion during the reaction. The energy of the emitted neutron depends on the center-of-mass velocity of the reacting initial ions and the direction of emission of the neutrons relative to the center of mass motion.

As the neutrons are uncharged, detection of theses particles are carried out either using a nuclear transformation reactions or neutron scattering events. Many types of detectors are used for neurotics, such as (1) Track recorder, which uses diffusion cloud or a bubble chambers, (2) Dosimeters uses foil activation, flow activation or moderator and activation, (3) Counters are also used for detecting the neutrons. It uses moderator along with $^{10}$BF$_3$ proportional counter or moderator along with $^{235}$U fission chamber, (4) Neutron spectrometers uses $^6$Li loaded scintillators along with fast photomultiplier tube, (5) Neutron imaging device uses pin hole camera along with a suitable detectors or a neutron streak camera with $^{235}$U coated photocathode.

These measurements can be divided mainly in three parts, such as total neutron yield (integrated over time), a time resolved neutron emission and energy spectra of neutrons.

Time –integrated measurements of neutron yield is based on the neutron activation of the foil samples. This technique is used to calibrate the neutron yield monitors during the plasma operation. Only a few activation reaction are suitable for 2.5 MeV neutrons. The $^{115}$In (n, n$^{''}$) $^{115}$In reaction is the most widely used with a short half-life of 4.5 h. For the measurement of 14 MeV neutron $^{63}$Cu (n, 2n) $^{62}$Cu reaction is the most sensitive and popular, which radiates γ emission with a half life of 9.7 min.

Time resolved measurements uses a $^{235}$U or $^{238}$U fission chambers surrounded by a polyethelene neutron moderator along with a suitable detector. Neutron streak camera is also used for getting information about the temporal evolution of neutrons in the fusion reaction, particularly in inertially confined fusion. Neutron streak camera uses the photocathode of $^{235}$U. Rests of the systems are same as the optical or X-ray streak camera.

Neutrons emitted from the D-D and D-T fusion reaction are not strictly monoenergetic (2.45 MeV and 14 MeV) but broadened due to the motions of the reacting ions. A careful measurement of the neutron energy spectrum allows determination of core ion temperature through the Doppler broadening. The broadening is relatively small ($\Delta E/E \leq 5\%$) and therefore a very good energy resolution is needed for the neutron spectrometer. For this purpose mainly a time of flight spectrometer has been used, which uses very fast scintillator along with a fast rise time photomultiplier tube. Total neutron yield from the fusion reaction as well as the number of elastically scattered deuteron and triton (charged particles of deuterium and tritium) can provide information about the areal density ($\rho R$), ($\rho$ is the density and R is the radius of the compressed fusion pellet) in the core of the fusion pellet. Elastically scattered charged fusion products are measured by counting the number of tracks in the CR-39 film.

Brief information about some of the important plasma diagnostics has been provided in this article. Still there are many more diagnostics for measuring the plasma parameters, but it is not possible to mention all of them here due to space limitation. However references are provided for further reading to get better insight about these as well as other advanced techniques in detail for the measurement of different plasma parameters.